\documentclass{article}
\pdfoutput=1 
\usepackage{spconf,amsmath,graphicx,bm,amssymb}
\usepackage{hyperref}
\usepackage{xcolor}
\usepackage[font=small,labelfont=bf]{caption}
\usepackage[square,sort&compress,comma,numbers]{natbib}
\title{Self-Supervised Physics-Based Deep Learning MRI Reconstruction Without Fully-Sampled Data}
 \name{\begin{tabular}{c}Burhaneddin Yaman$^{\star \dagger}$, Seyed Amir Hossein Hosseini$^{\star \dagger}$, Steen Moeller$^{\dagger}$, \\
 Jutta Ellermann$^{\dagger}$, K$\hat{\textrm{a}}$mil U$\check{\textrm{g}}$urbil$^{\dagger}$ and Mehmet Ak\c{c}akaya$^{\star \dagger}$
 \end{tabular}}

 \address{$^{\star}$ Electrical and Computer Engineering, University of Minnesota, Minneapolis, MN, USA \\
      $^{\dagger}$ Center for Magnetic Resonance Research, University of Minnesota, Minneapolis, MN, USA\\
            }

\begin{document}
\maketitle
\begin{abstract}
Deep learning (DL) has emerged as a tool for improving accelerated MRI reconstruction. A common strategy among DL methods is the physics-based approach, where a regularized iterative algorithm alternating between data consistency and a regularizer is unrolled for a finite number of iterations. This unrolled network is then trained end-to-end in a supervised manner, using fully-sampled data as ground truth for the network output. However, in a number of scenarios, it is difficult to obtain fully-sampled datasets, due to physiological constraints such as organ motion or physical constraints such as signal decay. In this work, we tackle this issue and propose a self-supervised learning strategy that enables physics-based DL reconstruction without fully-sampled data. Our approach is to divide the acquired sub-sampled points for each scan into training and validation subsets. During training, data consistency is enforced over the training subset, while the validation subset is used to define the loss function. Results show that the proposed self-supervised learning method successfully reconstructs images without fully-sampled data, performing similarly to the supervised approach that is trained with fully-sampled references. This has implications for physics-based inverse problem approaches for other settings, where fully-sampled data is not available or possible to acquire. 
\end{abstract}

\begin{keywords}
Self-supervised learning, accelerated imaging, parallel imaging, compressed sensing, deep learning, neural networks, supervised learning 
\end{keywords}
\section{Introduction}
\label{sec:intro}

Long acquisition times remain a limitation for MRI. In most clinical protocols, a form of accelerated imaging is utilized to improve spatio-temporal resolution. In these methods, data is sub-sampled, and then reconstructed using additional information. Parallel imaging \cite{Sense,Grappa} which utilizes redundancies between receiver coils, and compressed sensing \cite{lustig, ktFocus} which uses compressibility of images are two common approaches.

Recently, deep learning (DL) has gained interest as a means of improving accelerated MRI reconstruction \cite{DongLiang,sun2016deep,Hammernik,Hemant,mardani2017recurrent,JongChulYeeDLMagPhase,RAKI,JongChulYeLoss}. Several approaches have been proposed, including learning a mapping from zero-filled images to artifact-free images \cite{JongChulYeeDLMagPhase}, learning interpolation rules in k-space \cite{RAKI,JongChulYeLoss}, and a physics-based approach that utilizes the known forward model during reconstruction \cite{Hemant,Hammernik,sun2016deep,mardani2017recurrent}. The latter approach considers reconstruction as an inverse problem, including 
a data consistency term that involves the forward operator and a regularization term that is learned from training data. Such methods typically unroll an iterative algorithm for a pre-determined number of iterations to solve this inverse problem \cite{LeCun}. 
Specifics of these networks vary \cite{leslieYing}. For instance, in \cite{Hammernik}, data consistency used a gradient step and regularizer was a variational network, whereas in \cite{Hemant} conjugate gradient was used in data consistency and a residual network as regularizer.

To the best of our knowledge, most physics-based DL-MRI reconstructions to date use supervised learning, utilizing fully-sampled data as reference for training the network. However, in several practical settings, it is not possible to acquire fully-sampled data due to physiological constraints, e.g. high-resolution dynamic MRI, or due to systemic limitations, e.g. high-resolution single-shot diffusion MRI. %
Furthermore, accelerated imaging is often used to improve resolution feasible, which in turn makes it infeasible to acquire fully-sampled data at higher resolution due to scan time constraints. 

In this work, we propose a self-supervised learning approach that uses only the acquired sub-sampled k-space data to train physics-based DL-MRI reconstruction without fully-sampled reference data. Succinctly, our approach divides the acquired k-space points into two sets. The k-space points in the first set are used for data consistency in the network, while the k-space points in the second set are used to define the loss function. Results on knee MRI \cite{fastmri} show that our self-supervised training performs similar to conventional supervised training for the same network structure, while outperforming parallel imaging and compressed sensing.

\section{Materials and Methods}
\label{sec:meterialsandmethods}
\subsection{Physics-Based Deep Learning MRI Reconstruction}
\begin{figure}[t]
  	\begin{minipage}[b]{1.0\linewidth}
  		\centering
  		\centerline{\includegraphics[width=8.5 cm]{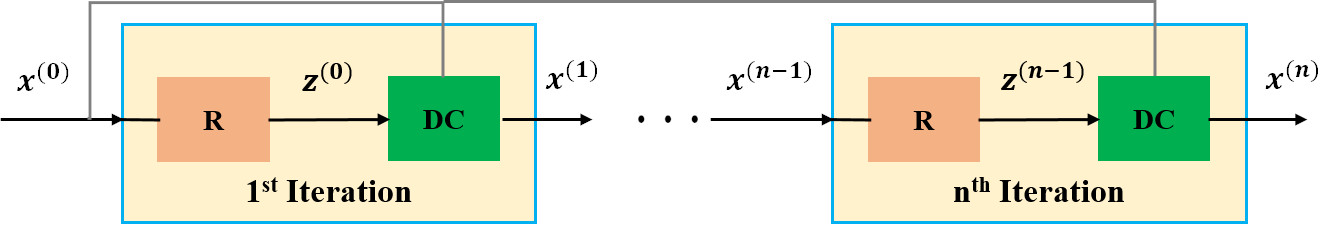}}
  	\end{minipage}

  	\caption{The unrolled neural network architecture for solving Equation (2), where each step consists of a regularization (R) and a data consistency (DC) unit. 
  	}
  	\label{fig:ResNet_architecture}
  	\vspace{-.3cm}
\end{figure}

Let $\mathbf{y}_\Omega$ be the acquired data in k-space, where $\Omega$ denotes the sub-sampling pattern of acquired locations, and $\mathbf{x}$ be the image to be recovered. The forward model is given as 
\begin{equation}\label{first_eq}
	{\bf y}_{\Omega} = {\bf E}_{\Omega} {\bf x} + {\bf n},
\end{equation}
where ${\bf E}_{\Omega}: {\mathbb C}^{M \times N} \to {\mathbb C}^P$ is the forward encoding operator, including a partial Fourier matrix and the sensitivities of the receiver coil array \cite{Sense}, and ${\bf n} \in {\mathbb C}^P$ is the measurement noise. Equation (\ref{first_eq}) is generally ill-posed and thus commonly solved using a regularized least squares problem \cite{lustig,Lingala} as follows
\begin{equation}\label{Eq:recons1}
\arg \min_{\bf x} \|\mathbf{y}_{\Omega}-\mathbf{E}_{\Omega}\mathbf{x}\|^2_2 + \cal{R}(\mathbf{x}),
\end{equation}
where first term enforces data consistency with acquired measurements, and $\cal{R}(\mathbf{\cdot})$ is a regularizer. There are several strategies to solve this optimization problem \cite{Fessler}, where methods alternate between enforcing data consistency with acquired data ${\bf y}_{\Omega}$ and a proximal operation involving $\cal{R}(\mathbf{\cdot})$. For instance, using variable-splitting and quadratic relaxation \cite{Fessler}, we have
\begin{subequations}
\begin{align}
& \mathbf{z}^{(i-1)} = \arg \min_{\bf z}\mu \lVert\mathbf{x}^{(i-1)}-\mathbf{z}\rVert_{2}^2 +\cal{R}(\mathbf{z})\label{Eq:recons3a}
\\
& \mathbf{x}^{(i)} = \arg \min_{\bf x}\|\mathbf{y}_{\Omega}-\mathbf{E}_{\Omega}\mathbf{x}\|^2_2 +\mu\lVert\mathbf{x}-\mathbf{z}^{(i-1)}\rVert_{2}^2\label{Eq:recons3b}
\end{align}
\end{subequations}
 where $\mathbf{z}^{(i)}$ is an intermediate variable and $\mathbf{x}^{(i)}$ is the desired image at iteration $i$. This algorithm can be unrolled for a fixed number of iterations \cite{LeCun}, as depicted in Figure \ref{fig:ResNet_architecture}.

 Physics-based DL-MRI methods train these types of unrolled algorithms end-to-end using fully-sampled training datasets \cite{leslieYing}. The sub-problem (\ref{Eq:recons3a}) is implemented by means of a neural network, while the data-consistency sub-problem (\ref{Eq:recons3b}) is solved via 
 \begin{equation}
\mathbf{x}^{(i)} = (\mathbf{E}_{\Omega}^H\mathbf{E}_{\Omega}+\mu \mathbf{I})^{-1}(\mathbf{E}_{\Omega}^H\mathbf{y}_{\Omega} + \mu\mathbf{z}^{(i-1)} ),
\end{equation}
where $\mathbf{I}$ is the identity matrix and $(\cdot)^H$ is the Hermitian operator. This can be solved via conjugate gradient to avoid matrix inversion \cite{Hemant}.
The unrolled network is then trained end-to-end, either by allowing different parameters for each iteration \cite{Hammernik} or by sharing all trainable parameters across iterations \cite{Hemant}.

\subsection{Supervised Unrolled Network Training}
In the supervised setting, SENSE-1 images generated from fully-sampled data are often utilized as ground truth for training. Let ${\bf x}_{\textrm{ref}}^i$ denote the ground truth image for subject $i$. Let $f({\bf y}_{\Omega}^i, {\bf E}_{\Omega}^i; {\bm \theta})$ denote the output of the unrolled network for sub-sampled k-space data ${\bf y}_{\Omega}^i$, and encoding matrix ${\bf E}_{\Omega}^i$ of subject $i$, where the network is parameterized by ${\bm \theta}$. These parameters are learned using
\begin{equation}
    \min_{\bm \theta} \frac1N \sum_{i=1}^{N} \mathcal{L}( {\bf x}_{\textrm{ref}}^i, \: f({\bf y}_{\Omega}^i, {\bf E}_{\Omega}^i; {\bm \theta})),
\end{equation}
where $N$ is the number of fully-sampled datasets in the training database, and $\mathcal{L}(\cdot, \cdot)$ is a loss function between the network output image and the reference image. Common choices for $\mathcal{L}(\cdot, \cdot)$ include $\ell_2$ norm, $\ell_1$ norm, mixed norms and perception-based loss \cite{HammernikLoss,Hammernik,Schlemper,MortezaLoss,QinUnrollnetwork,knoll2019deep}.

\subsection{Proposed Self-Supervised Network Training}
As discussed in Section \ref{sec:intro}, there are several scenarios where fully-sampled k-space data cannot be acquired. We propose to tackle this challenge by dividing the acquired sub-sampled data indices, $\Omega$ into two sets, $\Theta$ and $\Lambda$ as
\begin{equation}
    \Omega = \Theta \cup \Lambda,
\end{equation}
where $\Theta$ denotes a set of k-space locations that is used within the network during training, and $\Lambda$ denotes a set of k-space locations used in the loss function. In particular, in the absence of reference fully-sampled datasets, we minimize a loss function of the form
\begin{equation}
    \min_{\bm \theta} \frac1N \sum_{i=1}^{N} \mathcal{L}\Big({\bf y}_{\Lambda}^i, \: {\bf E}_{\Lambda}^i \big(f({\bf y}_{\Theta}^i, {\bf E}_{\Theta}^i; {\bm \theta}) \big) \Big).
\end{equation}
We note that the loss function is defined between the network output image and a vector of k-space points, in contrast to the supervised case, which traditionally has a loss function defined over the image domain. We hypothesize that by calculating the loss only on $\Lambda$ instead of the whole acquired data $\Omega$, the network will be better-suited to avoid over-fitting issues and generalize to future data.

\subsection{Implementation Details}
\label{sec:24}

Training was performed end-to-end by unrolling the algorithm in (\ref{Eq:recons3a})-(\ref{Eq:recons3b}) for 10 iterations, where each iteration consisted of regularization and data consistency units. The regularization convolutional neural network (CNN) employed a ResNet structure, consisting of a layer of input and output convolution layers and 15 residual blocks (RB) with skip connections that facilitate the information flow during training \cite{SNUCVLab}. Each RB comprised two convolutional layers, where the first layer is followed by a rectified linear unit (ReLU) and the second layer is followed by a constant multiplication layer \cite{SNUCVLab}. All layers had a kernel size of $3\times3$, 64 channels. The data consistency unit used a conjugate gradient approach, which itself was unrolled for 10 iterations \cite{Hemant}. Coil sensitivity maps in the encoding matrices were generated using ESPIRiT \cite{fastmri}. The network had a total of 592,129 trainable parameters. 

\begin{figure}[t]

  	\begin{minipage}[b]{1.0\linewidth}
  		\centering
  		\centerline{\includegraphics[width=8.5cm]{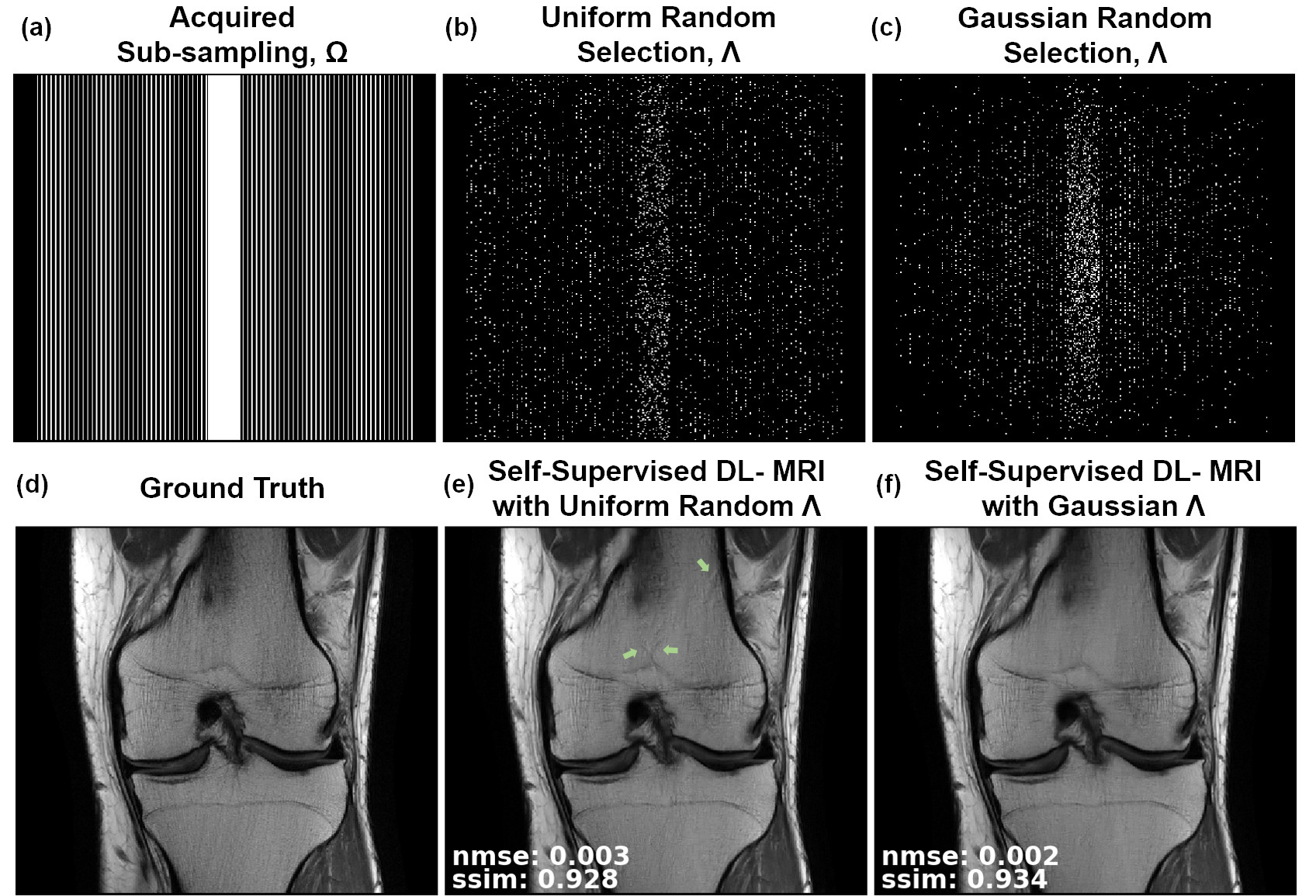}}
  	\end{minipage}

  	\caption{a) The acquired sub-sampling pattern $\Omega$; b) Uniform random and c) Gaussian random selection for subset $\Lambda$ over which loss is defined; d) Reference fully-sampled data; e) Self-supervised DL-MRI reconstruction with $\Lambda$ as in (b); f) with $\Lambda$ as in (c). Green arrow marks residual artifacts in uniform random selection of $\Lambda$. These are suppressed further with Gaussian random selection.
  	}
  	\label{fig:gauss_vs_unif}
 	\vspace{-.3cm}
\end{figure}

A normalized $\ell_1$ - $\ell_2$ loss, defined as
\begin{equation}
    {\cal L}({\bf u}, {\bf v}) = \frac{||{\bf u} - {\bf v}||_2}{||{\bf u}||_2} + \frac{||{\bf u} - {\bf v}||_1}{||{\bf u}||_1}
\end{equation}
was used for both the supervised and proposed self-supervised training. For the self-supervised setting, $\Theta$ was chosen as $\Omega / \Lambda.$ The networks were trained using an Adam optimizer with a learning rate of $10^{-3}$ by minimizing the respective loss function with a batch size of 1 over 100 epochs.

\subsection{Imaging Experiments}
\label{ssec:invivodatasets}
 Coronal proton density weighted knee MRI data from the New York University (NYU) fastMRI initiative database \cite{fastmri} was used for training and testing. Training data consisted of 300 slices from 10 patients. Each raw k-space data was of size $320 \times 368 \times 15$ where the first two dimensions are the matrix sizes and the last dimension is the number of coils. Testing was performed on 380 slices collected from 10 new subjects. 
 
 The fully sampled raw data were undersampled retrospectively using a uniform sub-sampling pattern provided by NYU with an acceleration rate of 4, and 24 lines as autocalibrated signal (ACS). The first set of experiments were designed to evaluate the choice of $\Lambda$ on the proposed self-supervised training. 
 Since $\Lambda$ is a retrospectively selected subset of $\Omega$ during reconstruction, its choice is not constrained by physical limitations, such as gradient switching. Thus, $\Lambda$ can be selected among all possible k-space locations in $\Omega.$ In particular, a uniform random selection of $\Lambda$, as well as a variable-density selection based on Gaussian weighting were investigated. Subsequently, the ratio $\rho = |\Lambda|/|\Omega|$ was varied among \{0.05, 0.1, 0.2, 0.3, 0.4\}, where $|\cdot|$ defines the cardinality of the index set.

 Following the study on the choice of $\Lambda,$ the proposed self-supervised learning method was compared with the conventional supervised learning approach, where the same network structure described in Section \ref{sec:24} were used for both approaches. Furthermore, the methods were compared to conjugate gradient SENSE (CG-SENSE) \cite{cgsense}, as well as a conventional compressed sensing approach that uses a total generalized variation (TGV) \cite{knoll_TGV} term for regularization in Equation (\ref{Eq:recons1}). Experimental results were quantitatively evaluated using normalized mean square error (NMSE) and structural similarity index (SSIM). 
\vspace{-.3cm}
\begin{figure}[b]
  	\begin{minipage}[b]{1.0\linewidth}
  		\centering
  		\centerline{\includegraphics[width=8.5cm]{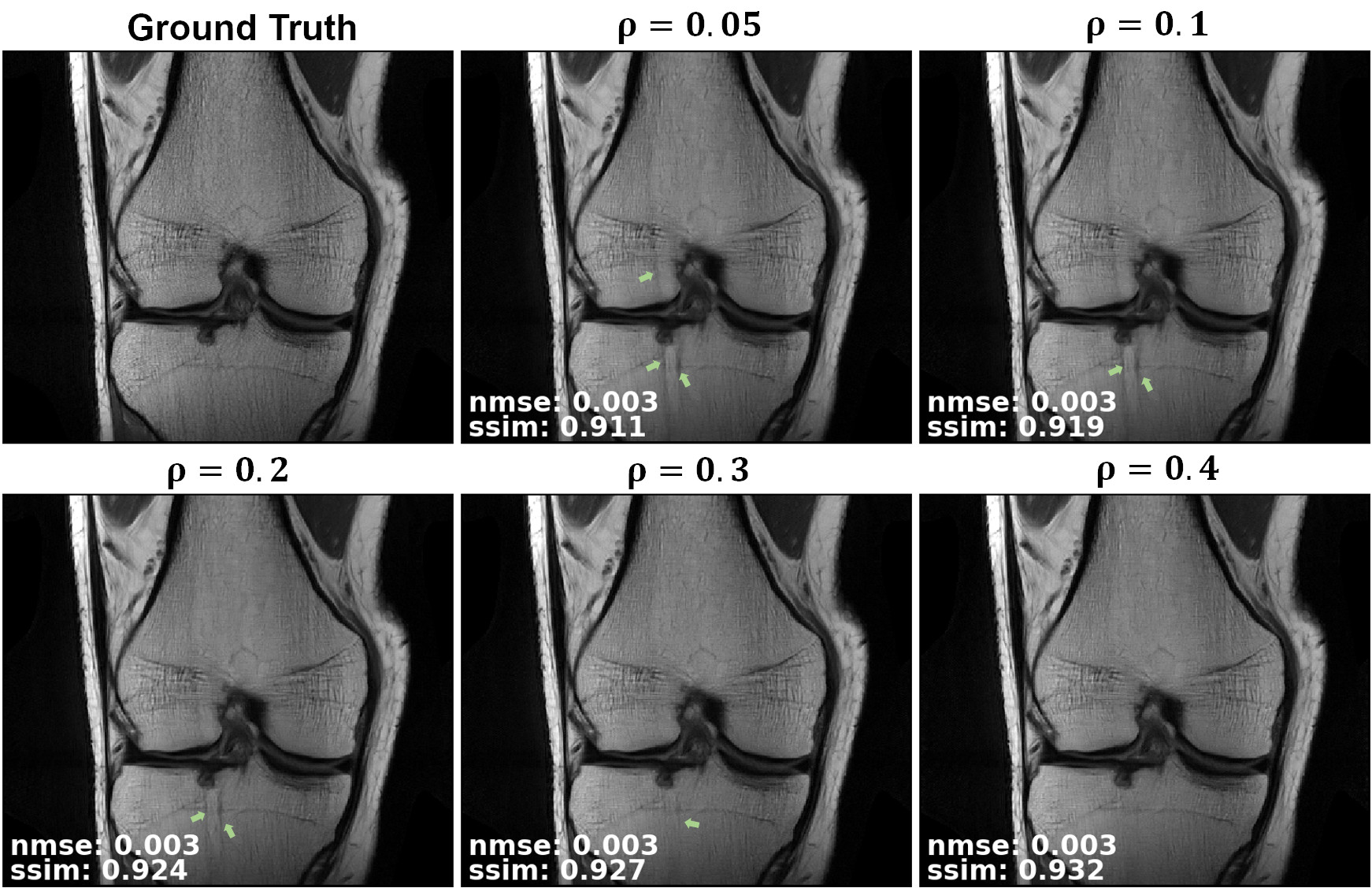}}
  	\end{minipage}

  	\caption{A test slice reconstructed using different ratios of $\rho = |\Lambda|/|\Omega|$, where $\Lambda$ is only used in the loss and $\Omega/\Lambda$ is only used in the data consistency unit of the network.
  	}
  	\label{fig:validation_percentages}
 	\vspace{-.3cm}
\end{figure}
\begin{figure*}[t]

  	\begin{minipage}[b]{1.0\linewidth}
  		\centering
  		\centerline{\includegraphics[width= 7 in]{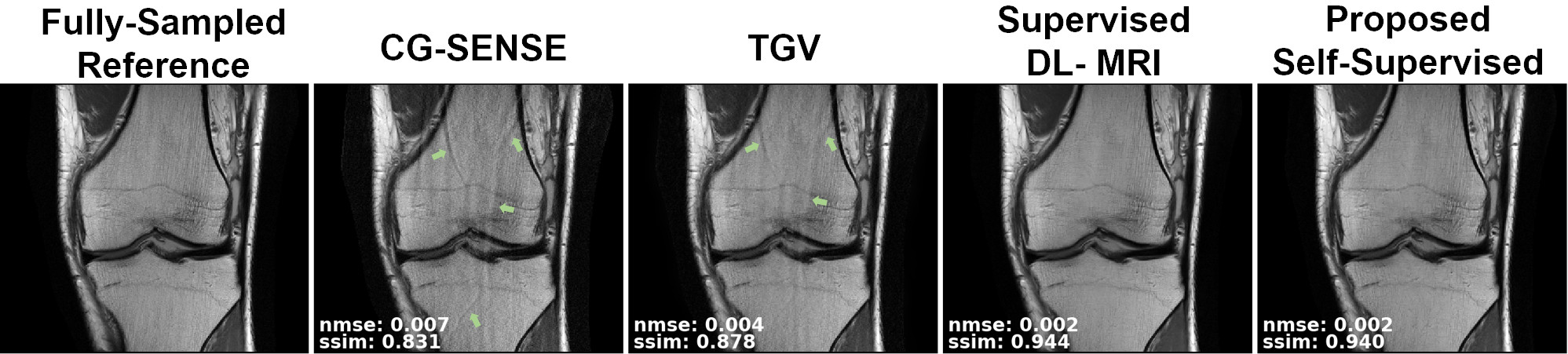}}
  	\end{minipage}

  	\caption{A slice from a proton-density knee dataset, depicting the reference fully-sampled image, and reconstructions using CG-SENSE, TGV, supervised DL-MRI trained on fully-sampled data and the proposed self-supervised DL-MRI trained on only sub-sampled data. The proposed training strategy leads to a DL-MRI reconstruction that removes artifacts successfully and perform similar to the supervised approach, while visibly outperforming CG-SENSE and TGV approaches.
  }
  	\label{fig:selfsupervised}
\end{figure*}
\section{Results}
Figure \ref{fig:gauss_vs_unif} shows the self-supervised network training with $\rho = 0.1$ for uniformly random and variable-density Gaussian selection of $\Lambda \subset \Omega.$ The green arrows show that visible residual artifacts remained in the reconstruction using uniform random selection of $\Lambda$. These artifacts were further suppressed by selecting a variable-density Gaussian subset as $\Lambda$. The corresponding quantitative evaluations, depicted in the figure, confirm these observations.

The effect of varying  $\rho \in$ \{0.05, 0.1, 0.2, 0.3, 0.4\} is shown in Figure \ref{fig:validation_percentages}. The residual artifacts, marked by green arrows, decrease with increasing ratio $\rho$ and disappear for $\rho \in$ \{0.3, 0.4\}. Quantitative results indicate that these two values have similar performance, with the latter showing slightly more visual improvement. Thus $\rho = 0.4$ was used for the rest of the study.

Figure \ref{fig:selfsupervised} displays the comparison among the proposed self-supervised, supervised, TGV and CG-SENSE methods. The green arrows on CG-SENSE and TGV results show visible residual artifacts. The supervised and the proposed self-supervised DL-MRI reconstruction approaches eliminate these artifacts, while performing closely with each other. The SSIM and NMSE values for this slice further confirms the visual assessments.
\begin{figure}[!b]

  	\begin{minipage}[b]{1.0\linewidth}
  		\centering
  		\centerline{\includegraphics[width=8.5cm]{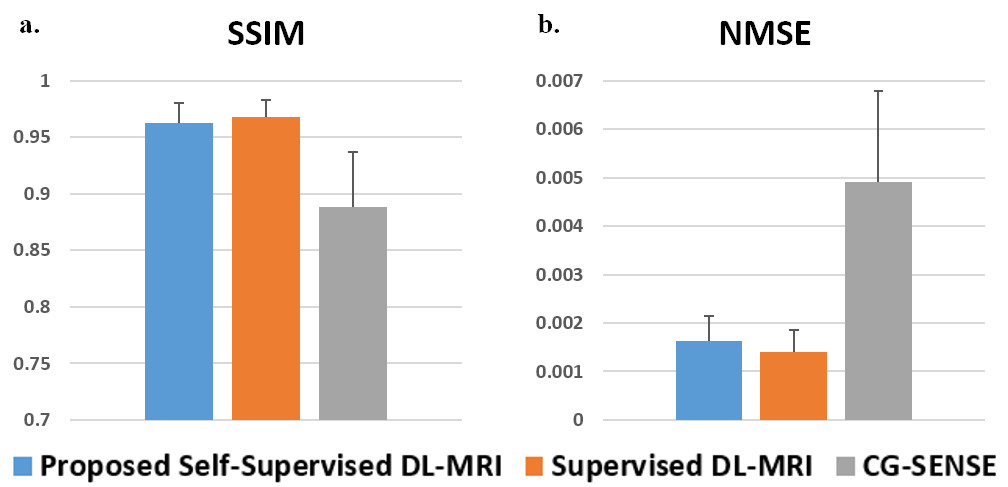}}
  	\end{minipage}

  	\caption{a) and b) shows the SSIM and NMSE values for the 380 slices tested. Supervised DL-MRI approach performs slightly better than the proposed self-supervised DL-MRI approach, while both methods readily outperform CG-SENSE quantitatively.}
  	\label{fig:BarPlot}
  	\vspace{-.3cm}
\end{figure}
Figure \ref{fig:BarPlot}a and b summarizes the mean and standard deviation of the NMSE and SSIM metrics over the 380 test slices using the proposed self-supervised DL-MRI, conventional supervised DL-MRI and CG-SENSE approaches. The two DL-MRI methods yield similar SSIM and NMSE values, even though the self-supervised approach does not use any fully-sampled datasets during training. Both methods outperform CG-SENSE.

\section{Discussion}
\label{sec:discussion}
We have proposed a self-supervised training method for physics-based DL-MRI reconstruction in the absence of fully-sampled data. The set of sub-sampled data indices $\Omega$ was divided into two sets $\Theta$ and $\Lambda,$ where the former was used during data consistency in the unrolled network, and the latter was utilized in the loss function. 
Results on fastMRI dataset indicate that our approach is successful in training a reconstruction algorithm that removes aliasing artifacts, achieving comparable performance to the conventional supervised learning approach that has access to fully-sampled data, while outperforming traditional compressed sensing and parallel imaging. 
The same neural network architecture was used for the self-supervised and supervised training.
While this is not the focus of this study, other choices of neural networks are  possible for further improvement. The effect of different choices of $\Lambda$ was also studied, where a variable-density selection with sufficient cardinality was favored.

In many scenarios, acquisition of fully-sampled data is challenging due to physiological and physical constraints. The lack of ground truth data hinders the utility of the supervised learning approaches in these scenarios. The proposed self-supervised approach relies only on available sub-sampled measurements. While we have focused on MRI reconstruction, the proposed approach naturally extends to other linear inverse problems, and has potential applications in other imaging modalities.
\section{Conclusion}
\label{sec:conclusion}
The proposed self-supervised learning strategy allows training of physics-based DL-MRI reconstruction without requiring fully-sampled data, while performing similar to conventional supervised learning approaches.
\section{acknowledgments}
\label{sec:acknowledgments}
This work was partially supported by NIH R00HL111410, NIH P41EB027061, NIH U01EB025144, NSF CAREER CCF-1651825.  Knee MRI data were obtained from the NYU fastMRI initiative database \cite{fastmri}. NYU fastMRI investigators provided data but did not participate in analysis or writing of this report. A listing of NYU fastMRI investigators, subject to updates, can be found at \url{fastmri.med.nyu.edu}.

\bibliographystyle{IEEEbib}
\bibliography{reference}
\end{document}